\documentclass{esannV2}
\usepackage[utf8]{inputenc}
\usepackage{graphicx}
\usepackage{amssymb,amsmath,array}
\usepackage{subcaption}
\usepackage{cite}
\usepackage{algorithm}
\usepackage{algpseudocode}
\usepackage{url}
\def\D{\mathcal{D}}
\begin{document}
%style file for ESANN manuscripts
\title{Improving Privacy Benefits of Redaction}

%***********************************************************************
% AUTHORS INFORMATION AREA
%***********************************************************************
\author{Vaibhav Gusain$^1$ and Douglas Leith$^1$
%
% Optional short acknowledgment: remove next line if non-needed
\thanks{This work was supported by Science Foundation Ireland grant 16/IA/4610.}
%
% DO NOT MODIFY THE FOLLOWING '\vspace' ARGUMENT
\vspace{.3cm}\\
%
% Addresses and institutions (remove "1- " in case of a single institution)
1- Trinity College Dublin - School of Computer science and statistics \\
Dublin - Ireland
%
% Remove the next three lines in case of a single institution
% \vspace{.1cm}\\
% 2- Trinity College Dublin - School of Computer science and statistics \\
% Dublin - Ireland
}
%***********************************************************************
% END OF AUTHORS INFORMATION AREA
%***********************************************************************

\maketitle

\begin{abstract}
We propose a novel redaction methodology that can be used to sanitize natural text data. Our new technique provides better privacy benefits than other state of the art techniques while maintaining lower redaction levels.
\end{abstract}

\section{Introduction}

Redaction is widely used to hide sensitive information from text. It is a process of replacing selected words with an uninformative \texttt{[MASK]} symbol and is typically carried out manually to redact Personal Identifiable information (PII) such as names addresses, etc \cite{medPII} from the text. 

% When the data is numerical then addition of appropriate noise e.g. Laplacian noise scaled proportionally to the differential privacy (DP) "sensitivity" of data, can be used to engorce differential privacy \cite{diffprivstat}. However adding noise 

Protecting privacy is especially challenging for text data because redacting specified words is rarely enough: the surrounding context can easily continue to reveal sensitive information \cite{browncontextmakeshard}. Even when the sensitive text is sanitized using word-level DP approaches \cite{Custext,santext}, it has been observed that sensitive attributes such as political views, medical condition, gender can still be leaked by the sanitized text dataset as whole \cite{wordlevelprivacylimits,gusainv}.

To limit the information revealed by the text data, redaction can be carried out such that a sensitive dataset $\D_0$ becomes indistinguishable from a safe dataset $\D_1$\footnote{Sensitive dataset $\D_0$ contains sentences which contains sensitive information such as a specific medical conditions etc and a safe dataset $\D_1$ is a public dataset that is suitable diverse and non-sensitive.}. The privacy gained in such cases depends on the percentage of words redacted from the sentences present in $\D_0$ and $\D_1$, and can be estimated by calculating the Renyi-divergence \cite{renyi_estimate} between the distributions of  $\D_0$ and $\D_1$ and converting this to an $(\epsilon,\delta)$ differential privacy estimate using concentrated differential privacy\cite{concentrated_dp}, for more details see \cite{gusainv}.This approach although promising, can require almost 80\% of the words from the input text to be redacted in order to achieve a reasonable level of privacy. At that point there is almost no information left in the sentence and it does not have much utility left.

% Figure-\ref{comparing_embeddings} illustrates the effect of redaction on the renyi-divergence for medal dataset \cite{medal}. We observe that as the redaction is increased the divergence between the $\D_0$ and $\D_1$ decreases. 

In this paper we propose a novel redaction methodology which builds upon the work of the authors of \cite{gusainv}. We show that our approach provides better privacy guarantees while requiring much lower redaction levels. For example, achieve $\epsilon =0.01$ by only redacting 20-30\% of the words, which is the considerable improvement over the current state-of-the art methods. We also provide an open-source implementation of a KL-divergence loss (in PyTorch) which calculates KL-divergence from the sentence embeddings.

% Our contributions in this paper are: \texttt{1.)} We propose a novel approach to do redaction such that provides better privacy guarantees while maintaining low redaction levels compared to other redaction approaches. \texttt{2.)} We provide an open-source implementation of a KL-divergence loss (in PyTorch) which calculates KL-divergence from the sentence embeddings. 
% \texttt{3.)} We show that using fine-tuned transformer embeddings to calculate Renyi-divergence provides better privacy guarantee. 

\section{Related Work}
% There are two major approaches used to santize input training data to gain privacy benefits:

% \subsection{Redaction}
\texttt{Text redaction.} Most of the current approaches manually redact PII from the input text ~\cite{detectstudent,secretdetector1}. These approaches focus more on hiding user specific sensitive information from the input text rather than on information that is revealed by the input text as a whole. Recent research by \cite{gusainv}, proposed an approach to limit information revealed by the text using redaction. They use a logistic regression model to rank and redact the words. They observe an increase in privacy benefits as redaction levels are increased. 

\texttt{Word level DP} Another approach to santize input text is to use Word-level DP. These approaches map an individual word to a vector embedding, add noise and then either map back to a new word or use the noisy embedding directly. See e.g. \cite{santext,Custext}. However sensitive attributes such as political views, medical condition, gender can still be leaked by the sanitized sentences \cite{wordlevelprivacylimits,gusainv}.

\texttt{Renyi-Divergence} Divergence is widely used to calculate the distance between two probability distributions \cite{renyi_estimate}. Renyi-Divergence of order $\alpha$ between two probability distributions $P_0$ and $P_1$ on sample space $Y$ is \cite{renyi_estimate}:

\begin{align}\label{eq1} 
D_{\alpha}(P_0||P_1) = \frac{1}{\alpha-1}\log\int_YP_0(x)^\alpha P_1(x)^{1-\alpha}dx
% \int_YP_0(dx)\log\frac{P_0(dy)}{P_1(dy)}
\end{align}
and similarly for $D_{\alpha}(P_1||P_0)$. When $\alpha =1$ the Renyi-divergence equals the KL-divergence ~\cite{renyi_estimate}.  
% For textual datasets $\D_0$ and $\D_1$ probability distribution $P_0$ and $P_1$ can be generated using a KNN \cite{knn_paper} estimator on the embeddings of the sentences present in $\D_0$ and $\D_1$. Divergence is then calculated using $P_0$ and $P_1$, for information see \cite{mauve,gusainv}. 
Divergence can be converted to a $(\epsilon,\delta)$ differential privacy guarantee using concentrated differential privacy. Due to lack of space we do not include the details here, but refer the reader to \cite{gusainv,concentrated_dp} for complete proofs.

\section{Overall Architecture}
% \begin{figure}[htb]
%   \centering
%   \includegraphics[width=0.5\textwidth]{overall_approach.png}
  
%   \caption{New redaction technique \textbf{Change this to EPS}}
% \label{newredaction}
% \end{figure}
\begin{figure*}[tb]%[t!]
\centering
     \begin{subfigure}{0.30\textwidth}
        \centering
            \includegraphics[width=\textwidth]{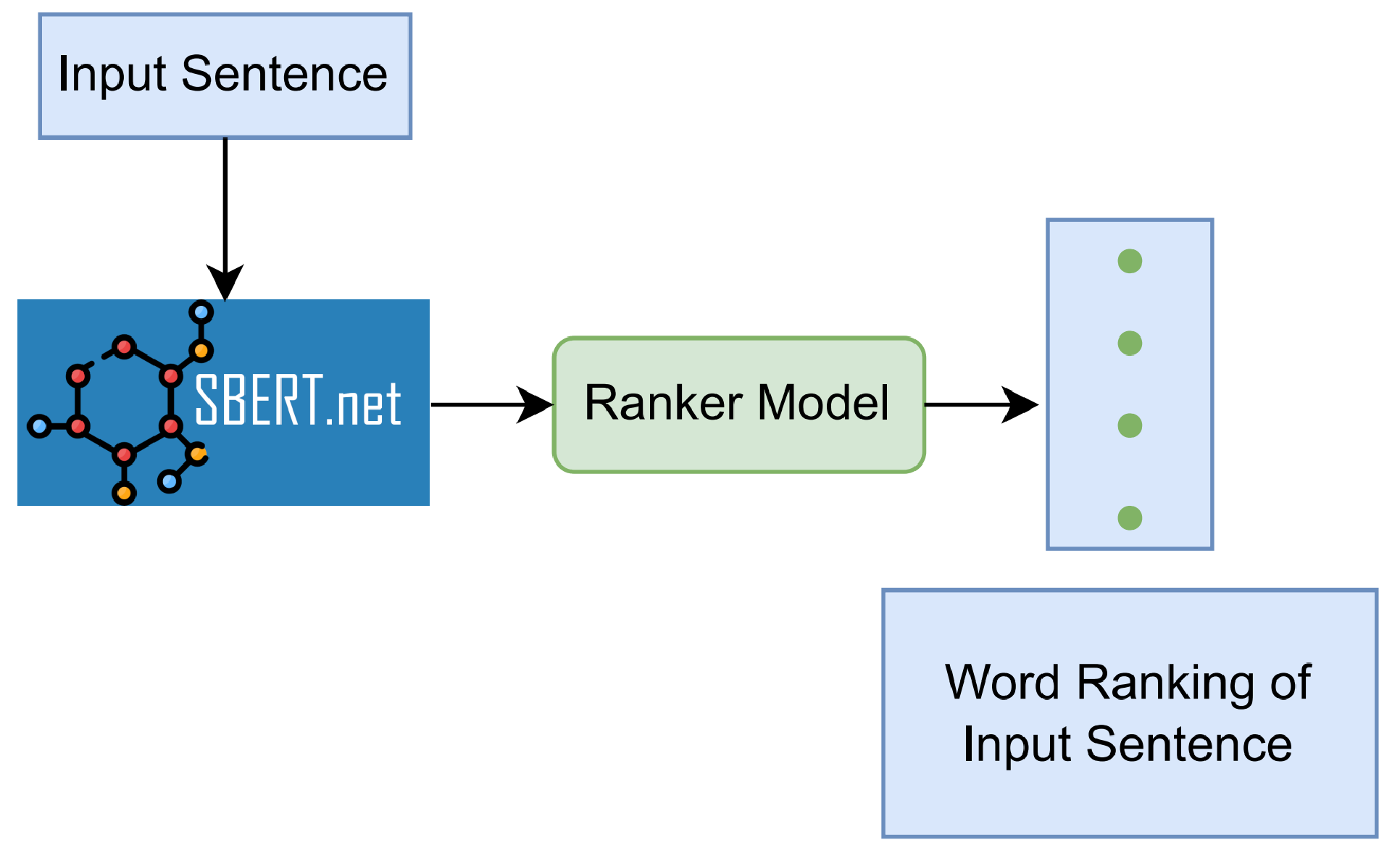}
            % \caption{New redaction technique \textbf{Change this to EPS}}
            \caption{}
            \label{newredaction}
    \end{subfigure}
    \hfill
    \begin{subfigure}{0.30\textwidth}
        \centering
            \includegraphics[width=\textwidth]{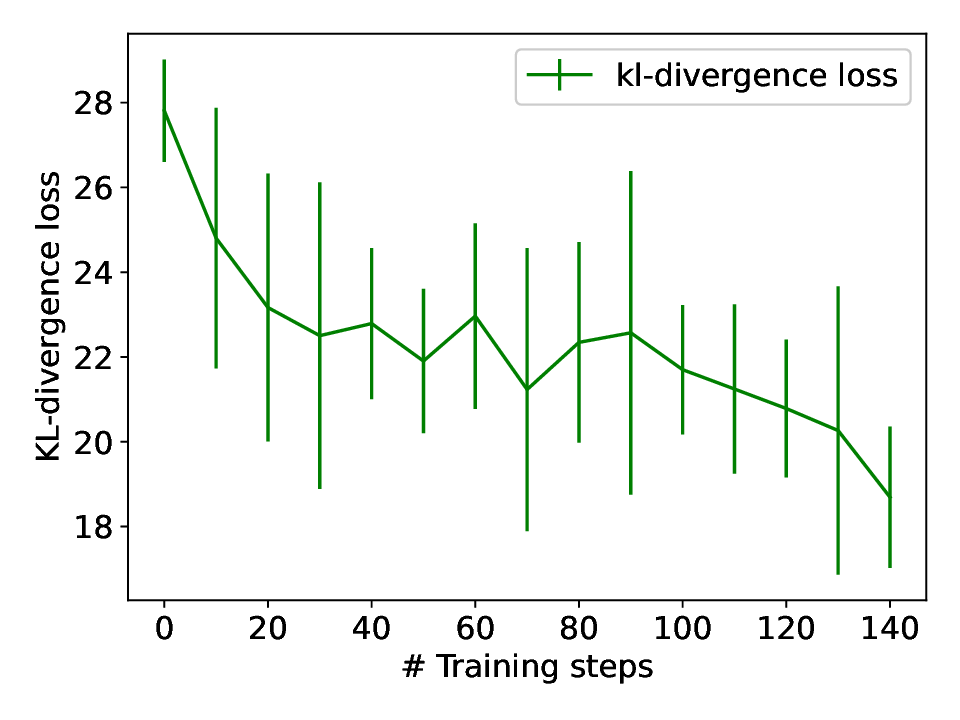}

            \caption{}
            \label{kldivloss}
    \end{subfigure}
     % \hfill
     \begin{subfigure}[b]{0.30\textwidth}
         {\includegraphics[width=\textwidth]{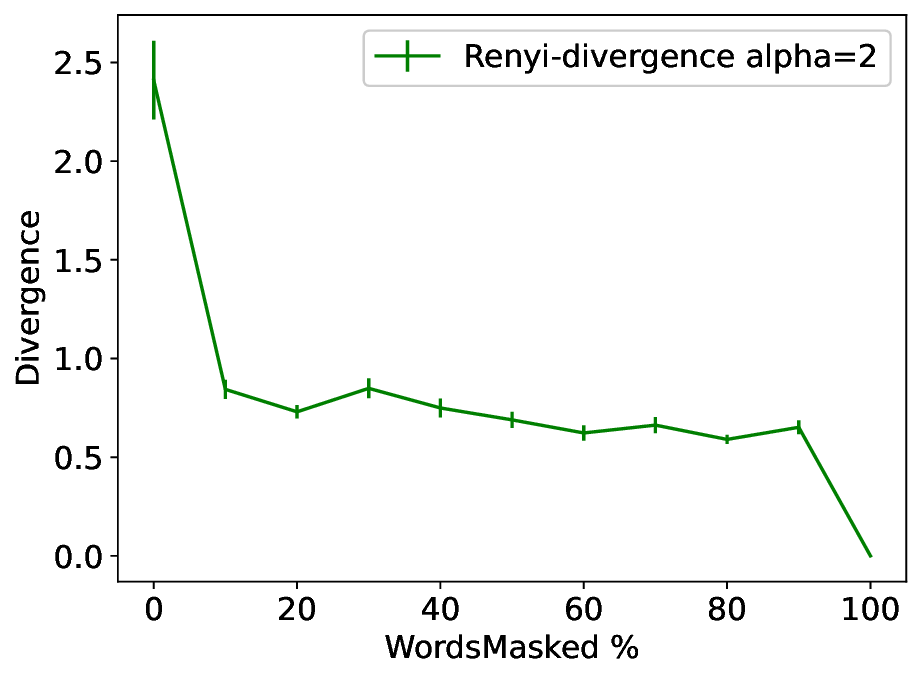}}%
         \caption{}
         \label{comparing_embeddings}
     \end{subfigure}
% \newline
\caption{\texttt{a:}Architecture of the new redaction technique. \texttt{b:} Average KL-divergence loss vs number of training steps on Medal dataset; Average loss is calculated at every 10 steps. \texttt{c:} Measured Renyi divergence for $(\alpha = 2)$ vs redaction level for Medal dataset; smarter-redaction is used, see Section \ref{redact_sec}. }
% \vspace*{5mm}
\label{distances_between_weights}
\end{figure*}
% Figure-\ref{newredaction} shows the new smarter redaction approach we used in this paper. In the following section we will delve into more details about each of the components.
% \subsection{Model Architecture}
Our method consists of two modules as shown in Figure-\ref{newredaction}:

\texttt{1.)Sentence transformer}. This is a sentence transformer model \cite{sentencebert} that is responsible for generating contextual word-embeddings for an input sentence $x_i$. 

\texttt{2.)Ranker Model}. This is a neural network consisting of 4 Linear layers, the first three layers use a tanh activation  while the final layer uses a sigmoid activation. This is responsible for ranking the words from an input sentence. It takes the embedding of a word present in the sentence $x_i$ as input and outputs the corresponding ranking for the word. 

To redact words from an input sentence, we first generate the embedding of each word in the sentence using the sentence transformer. The word embeddings are then sent to the ranker which generates ranking for each individual word. Words with lowest the K\% of the rank are redacted from the input sentence.

% It takes word embeddings of a sentence $x_i$ as an input, and outputs a ranking vector $r_i$ of length $W$ ($W$=number of words in the sentence). Each element of the vector $r_i$ provides a ranking to the corresponding word in the input sentence $x_i$. Words with lowest K\% rankings are redacted from the input sentence.

% Ranker is a neural network consisting of four linear layers each layer having a tanh activation at the output. 
% Each linear layer applies an affine transformation $y = x\cdot W^T+ b$, to an input vector $x$ to get an output vector $y$, where $W$ and $b$ are the weight and the bias vectors of a linear layer. Since this is a matrix multiplication, size of the output vector depends on the size of the input vectors $x$, $W$ and $b$, i.e. if size of the input vector $x$ is $( \ast ,H_{in})$, $W$ is $(H_{out},H_{in})$, $b$ is $(H_{out})$ then shape of output vector $y$ becomes $( \ast ,H_{out})$. Using 4 linear layers the ranker model converts an input embeddings of shape $(b,W,H_{in})$ to a ranking vector of shape $(b,W,1)$. Where $b$ is the batch-size of the input, $W$ \footnote{shorter sentences in the batch are padded such that each sentence has $W$ number of words.}  is the number of words in the longest sentence of the batch and $H_{in}$ is the length of each word embedding vector.
 
% For more information about the training procedure please go to Section-\ref{training_algo}. 

\section{Training Overview}\label{training_algo}
In this section we briefly discuss the training strategy used to train the ranker model. 

\subsection{Training the Ranker}

% Algorithm-\ref{alg:trainranker} illustrates the pseudo code used to train the ranker. Ranker takes word-embeddings of the sentences as the input and outputs the rank of the words that needs to be redacted.

During a training step a batch of sentences $db_0$ and $db_1$ is selected from separate held-out training datasets $D_0$ and $D_1$. Shorter sentences in a batch are padded with a \texttt{[pad]} token to make every sentence have same number of words $W$\footnote{\label{footn}$W$ is the number of words in the longest sentence from $db_0$ and $db_1$.  During our training we set K=10 and T=100.}. Using a sentence transformer word embeddings $e_0$ and $e_1$ for the padded sentences in $db_0$ and $db_1$ are generated. The ranker is then used to generate ranking vectors $r_0$ and $r_1$ from $e_0$ and $e_1$ respectively. The ranking vector for each sentence is updated such that lowest K\%$^{\ref{footn}}$
%\footnote{During our training we set K=10.} 
of the word rankings are set to zero and the rest are set to one. 
% Since boolean operation is not differentiable
To make this operation differentiable, we find the $K^{th}$ smallest rank $k_r$ from the ranking vector and subtract $k_r$ from each value in the ranking vector. The resulting vector is then multiplied with a hyper-parameter "T"$^{\ref{footn}}$
%\footnote{ During the training we set T=100.}
and passed through a sigmoid function to get an updated rank vector where the lowest K\% of the word-ranks are set to zero and the rest are set to one. The updated rank vector is multiplied with word embeddings $e_0$ and $e_1$ to get weighted embeddings $ue_0$ and $ue_1$. Sentence embeddings are generated from $ue_0$ and $ue_1$ by taking the average over the word embeddings of the sentence, which are then used to calculate the KL-divergence loss for the batch (see Section \ref{loss_custom})\footnote{We use KL divergence rather than Renyi divergence because the estimator is differentiable.}. An outline of the Algorithm is present in the appendix.\footnote{\url{https://github.com/vaibhav0195/density_estimation_code/blob/main/appendix.pdf}}

\subsection{KL-divergence Loss}\label{loss_custom}

For natural language datasets the embeddings from a sentence transformer can be used to estimate the probability distributions of two datasets, which can then be used to estimate divergence between the two, see for example \cite{mauve,gusainv}. We used the implementation provided by ~\cite{gusainv}, and modified it to create a custom loss function to calculate the KL-divergence between two natural language datasets using PyTorch. To the best of our knowledge there is no open-source implementation in PyTorch to calculate the KL-divergence between two sentence embeddings. The ranker model is trained using this custom loss function.

We can expect the ranker to randomly rank words when the training starts thus resulting in higher loss value. But as the training progresses, ranker will be optimized to rank the words such that redacting low K\% of the input words will result in lower divergence value between the two datasets. Figure-\ref{kldivloss} shows an example of the average loss vs the number of training steps. We observe that as the training progresses the average loss value is decreased. For more details about the hyper-parameters we refer the reader to an appendix.\footnote{\url{https://github.com/vaibhav0195/density_estimation_code}}
% Note. There  KL divergence
\section{Experiments}
\begin{figure*}[tb]%[t!]
\centering
     \begin{subfigure}{0.24\textwidth}
        \centering
            \includegraphics[width=\textwidth]{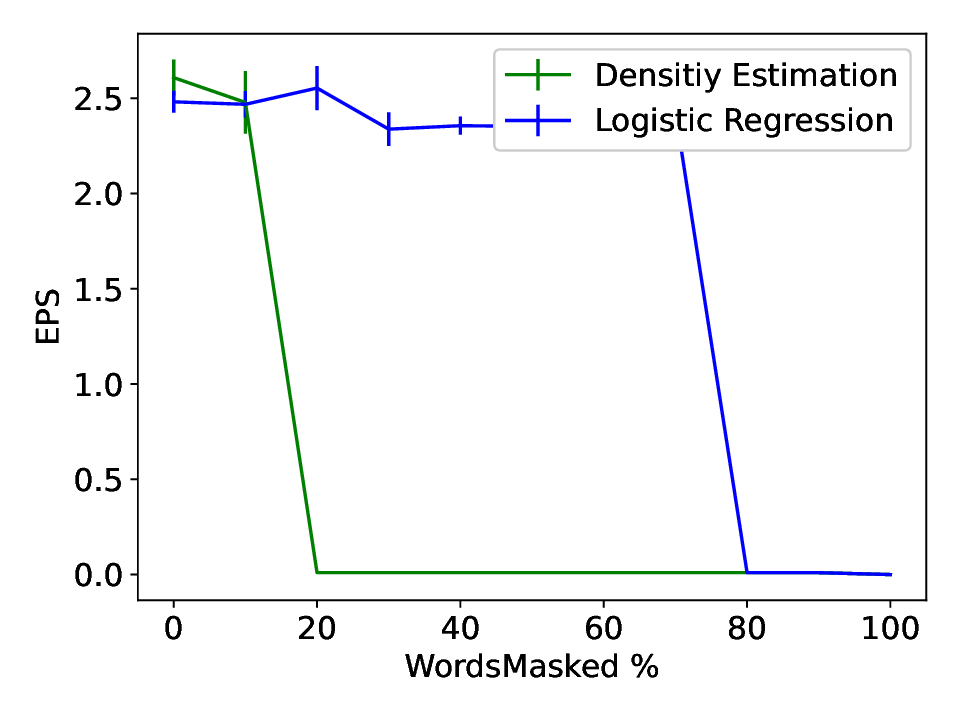}
            % \caption{New redaction technique \textbf{Change this to EPS}}
            \caption{Medal Dataset}
            \label{medal_dataset_comparison}
    \end{subfigure}
    \hfill
    \begin{subfigure}{0.24\textwidth}
        \centering
            \includegraphics[width=\textwidth]{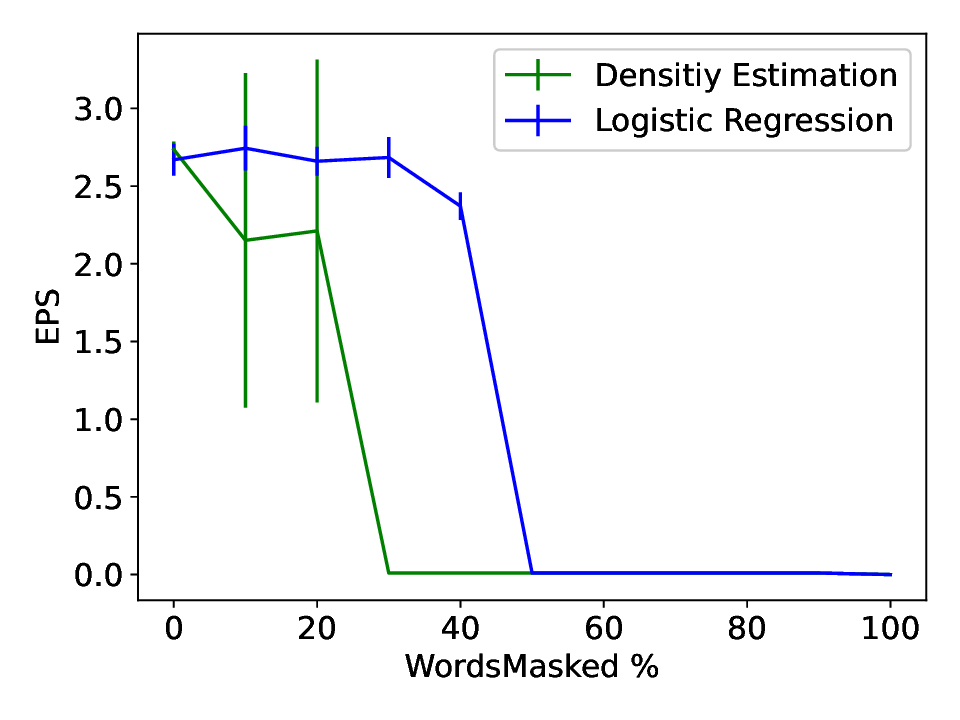}

            \caption{Political Dataset}
            \label{political_dataset_comparison}
    \end{subfigure}
     % \hfill
     \begin{subfigure}[b]{0.24\textwidth}
         {\includegraphics[width=\textwidth]{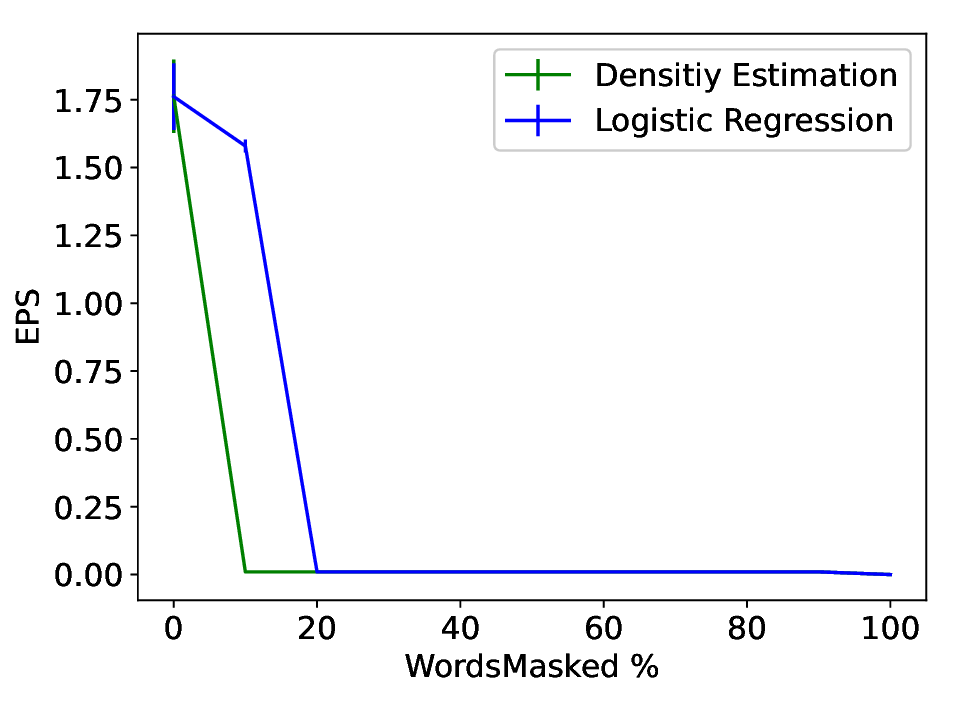}}%
         \caption{Amazon Dataset}
         \label{amazon_dataset_comparison}
     \end{subfigure}
     \begin{subfigure}[b]{0.24\textwidth}
         {\includegraphics[width=\textwidth]{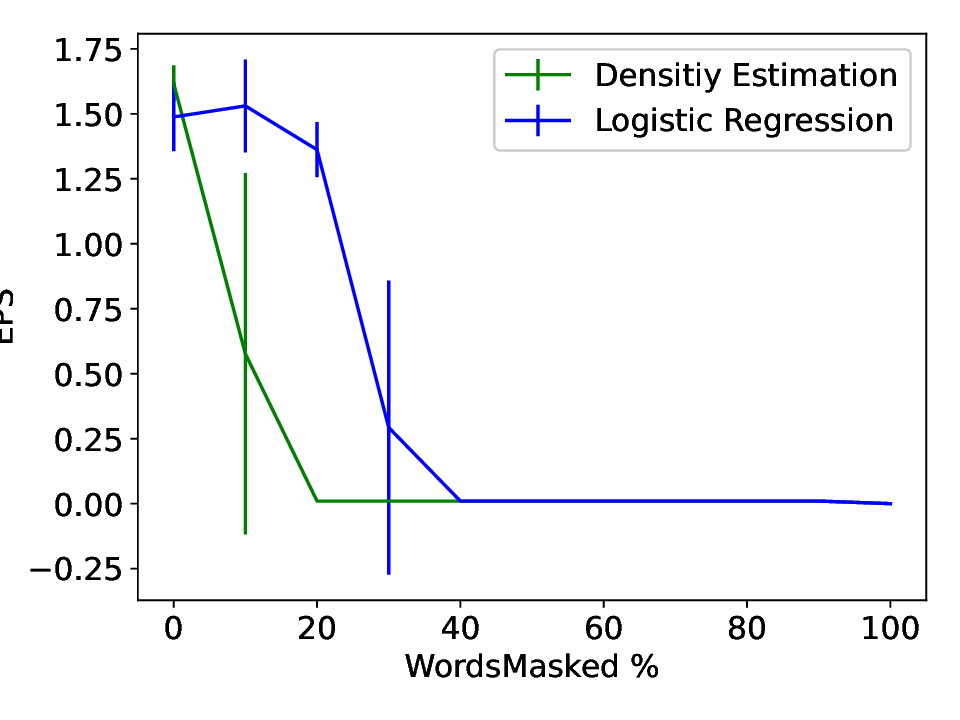}}%
         \caption{Reddit Dataset}
         \label{reddit dataset comparison}
     \end{subfigure}
% \newline
\caption{Measured $\epsilon$ between redacted sensitive and safe datasets vs redaction level;Comparison with various redaction strategy.; For our experiment we use $\delta = \frac{1}{n}$ \cite{gusainv}; n = number of input sentences; }
% \vspace*{5mm}
\label{results_comparison}
\end{figure*}

\subsection{Datasets}

We compared the performance on four datasets : Medal dataset \cite{medal}, Political dataset \cite{politicsdata}, Amazon dataset \cite{gusainv}, Reddit dataset. Sensitive and safe datasets were created from these datasets as explained in \cite{gusainv}. Each dataset contains almost $10000$ sentences each in the safe and sensitive dataset. The validation set for medal reddit and amazon dataset contained $2000$ sentences each in the safe and sensitive dataset. For more details regarding the datasets we refer reader to appendix.

% \subsection{Divergence Estimator}

% Note. We train our ranker using a KL divergence but present our results in the form of $(\epsilon,\delta)$ differential privacy which are derived from Renyi-divergence. 
% which is a special case of Renyi-divergence-($\alpha=1$) as explained earlier. 

% To calculate $(\epsilon,\delta)$ differential privacy for a given redaction percentage, we calculate renyi-divergence for various $\alpha$ values. We then find a line
% \texttt{you use KL divergence for training, but Renyi for the privacy evaluation, you need to comment on that difference}

\subsection{Redaction} \label{redact_sec}

Redaction was carried out on a separate validation set as the percentage of words redacted is varied. A fine-tuned sentence transformer was used to generate embeddings for the redacted sentences. The embeddings were used to estimate the Renyi-divergence. Renyi-divergence was estimated using the estimator provided in \cite{gusainv}, which was then converted to an $(\epsilon,\delta)$ differential privacy guarantee using concentrated differential privacy.

We compared our redaction approach against the smarter-redaction approach introduced in \cite{gusainv}. To redact words using smarter redaction, we trained a logistic regression model on TFIDF features of the training data. The trained model's weights were used to rank the words present in the sentence. Top $K$ percent of these words were then redacted. Note that we did not compare against random redaction as it does not provide any benefits over smarter redaction \cite{gusainv}.

% Redaction was carried out on a separate validation set using both logistic regression and our ranker model as the level of redaction i.e. percentage of words being redacted is varied. Sentence embeddings for the redacted sentences was generated using a fine-tuned sentence transformer.

% against two redaction approaches : \texttt{1.) Random Redaction:} Randomly redacting words from both sensitive and safe dataset and \texttt{2.) Redacting using logistic regression:} A logistic regression model is trained on TFIDF features of the training data that used to train the ranker. The model is then used to rank the words present in a sentence, top $p$ percent of these rank words are then redacted. For information see \cite{gusainv}.

\subsection{Results}

Figure-\ref{results_comparison} illustrates the privacy benefits of our approach compared to the smarter redaction approach introduced in \cite{gusainv}. Sentences ranked by our ranker achieve lower $\epsilon$ values for the same redaction percentage. We get $(\epsilon)$ values which are close to zero by only redacting 20-30\% of words, whereas the approach in \cite{gusainv} needs to redact almost 80\% of the input words to get similar privacy benefits.

We observe that our new ranker redacts words so as to quickly remove sensitive information early from the text which results in lower $\epsilon$ at lower redaction levels. E.g. consider the input sentence from Medal dataset \texttt{organ failure tone and ventral pallidum cell injury}, the important words to be redacted are "tone" and "ventral" cell as they reveal more about the type of injury. The logistic regression redacts the words "cell" and "failure", whereas our new ranker redacts "tone" and "ventral".

\section{Conclusion}

In this paper we propose a new loss function which can be used to train a neural network to redact words efficiently from sensitive text to gain privacy benefits. 

In addition to the neural network presented here we experimented with various other redaction approaches as well:- 1.) Redacting words using KNN- redacting words from the input sentence such that clusters between the two distribution $\D_0$ and $\D_1$ overlap quicker thus resulting in lower divergence values and 2.) Reinforcement learning- training a transformer model to generate domain and adaptive masking using reinforcement learning as explained in \cite{kang2020neural}, which can then be used to mask the words from the input sentence. We observe no significant improvement over the logistic regression approach. introduced in \cite{gusainv}.

In this work we only provide an estimate of Renyi-divergence and hence can not provide a theoretical privacy guarantee. However as pointed by \cite{gusainv} use of an estimate seems unavoidable since the true divergence cannot be calculated for realistic text data.  It is worth noting that there is a growing trend towards using empirical analysis in differential privacy, e.g. for auditing and for investigating the impact of changes in the threat model~ ~\cite{usenixaudit,audit_o1}.

\begin{footnotesize}
\bibliographystyle{plain} % We choose the "plain" reference style
% \bibliography{refs}
\bibliography{esannV2}

\end{footnotesize}

% ****************************************************************************
% END OF BIBLIOGRAPHY AREA
% ****************************************************************************

\end{document}

%style file for ESANN manuscripts
\title{Improving Privacy Benefits of Redaction-Appendix}

%***********************************************************************
% AUTHORS INFORMATION AREA
%***********************************************************************
\author{Vaibhav Gusain$^1$ and Douglas Leith$^1$
%
% Optional short acknowledgment: remove next line if non-needed
%
% DO NOT MODIFY THE FOLLOWING '\vspace' ARGUMENT
\vspace{.3cm}\\
%
% Addresses and institutions (remove "1- " in case of a single institution)
1- Trinity College Dublin - School of Computer science and statistics \\
Dublin - Ireland
%
% Remove the next three lines in case of a single institution
% \vspace{.1cm}\\
% 2- Trinity College Dublin - School of Computer science and statistics \\
% Dublin - Ireland
}
%***********************************************************************
% END OF AUTHORS INFORMATION AREA
%***********************************************************************

\maketitle

\section{Privacy Defination $(\epsilon,\delta)$}

% A dataset $\D$ is a collection of items.  Each item is a sequence \(x = (x_1,....., x_{|x|})\) of words \(x_t\) belonging to a fixed vocabulary and with length $|x|\le N$, $N$ being the maximum admissible length.  A redaction policy $\pi_p(x)$ maps sequence $x$ to a new sequence where some words have been redacted i.e. replaced by an uninformative mask token {\tt MASK}.  We will assume that every redaction policy is parameterised by a parameter $p$ taking a value between 0 and 1 such that when $p=0$ then no words are redacted, when $ p=1$ then every word is redacted.   For example, the uniform random $\pi_{rand,p}(x)$ redaction policy redacts each word in sequence $x$ with probability $p$.   Alternatively, we might rank the words in our vocabulary by their sensitivity and redact the top $p$ fraction of these.

Considering two datasets $\D_0$ and $\D_1$. Each item $x$ in a dataset is a random draw from a probability distribution $P(x)$ over sequences of words.  After redaction, each element $x$ is mapped to a new sequence $redact(x)$ and the redacted dataset becomes a sample from probability distribution $redact(P)$.   The distance between two redacted datasets $redact(\D_0)$ and $redact(\D_1)$ can be measured by the smallest value of $\epsilon\ge 0$ such that $\tilde{P}_0(y) \le e^\epsilon \tilde{P}_1(y) + \delta $ and $\tilde{P}_1(y) \le e^\epsilon \tilde{P}_0(y) + \delta $ where $\tilde{P}_0:=redact(P_0)$ is the probability distribution over token sequences in dataset $redact(\D_0)$, $\tilde{P}_1=redact(P_1)$ in dataset $redact(\D_1)$ and $y$ is any redacted sequence of words with length $|y|\le N$.  

This distance measure is similar to that used in $(\epsilon,\delta)$-differential privacy but with the difference that the set of neighbouring databases now consists of the single database $redact(\D_0)$  rather than all databases differing from $redact(\D_1)$ by a single element.  When $\epsilon,\delta$ are sufficiently small, the publication of private dataset $redact(\D_1)$ then only provides an attacker with limited new information over and above that already available from the public dataset $\D_0$.  That is, we gain privacy in the sense of \emph{indistinguishability} between the $redact(\D_0)$ and $redact(\D_1)$ datasets. It will prove convenient to work in terms of the Renyi-divergence $D_{\alpha}(\tilde{P}_0||\tilde{P}_1)$ to calculate the distance between the datasets. We then convert this to an $(\epsilon,\delta)$-privacy guarantee using concentrated differential privacy.  See \cite{gusainv} for more details.

% \subsection{}
\section{Outline for algorithm}
Algorithm-\ref{alg:trainranker} shows a pseudocode for training the ranker.
\begin{algorithm}[tb]
\caption{Train Ranker. $D_0$ represents the sensitive dataset. $D_1$ represents the safe dataset. sent\_trans is the sentence transformer fine-tuned on the training data. lossfn is the custom KL-divergence loss explained in Section-\ref{loss_custom}. bsz is the batchsize. sgd is stochastic gradient descent algorithm which is used to update the weights.}\label{alg:trainranker}
\begin{algorithmic}
% \small
\Function{\textbf{: train\_ranker}}{$D_0,D_1,sent\_trans,model,lossfn,bsz,sgd$}

\State {$k$ $\gets$ {$0$} ; $N$ $\gets$ {$len(D_0)$}}
\State {$randomize(D_0)$; $randomize(D_1)$}

\While{$k \leq N$}
    \State $db_0 \gets D_0[k:k+bsz]$ ; $db_1 \gets D_1[k:k+bsz]$
    \State $e_0 \gets sent\_trans(db_0)$ ; $e_1 \gets sent\_trans(db_1)$
    % \State $e_1 \gets E_1[k:k+bsz]$
    \State $r_0 \gets model(e_0)$ ; $r_1 \gets model(e_1)$
    \State $ue_0 \gets e_0 \cdot r_0$ ; $ue_1 \gets e_1 \cdot r_1$
    \State $loss \gets lossfn(ue_0, ue_1)$
    \State $model \gets sgd(model,loss)$
    \State $k \gets k+bsz$
\EndWhile
\State {\textbf{return} {$model$}}
\EndFunction
\end{algorithmic}
\end{algorithm}

% \begin{figure}[t]
%   \centering
%   \includegraphics[width=0.3\textwidth]{sent_bert_training_effect/comparison_embs.eps}
  
%   \caption{ Estimated $\epsilon$ and attack accuracy values for Medal dataset; We compare embeddings of a pretrained transformer to a fine-tuned transformer. }
% \label{comparing_embeddings}
% % \vspace*{5mm}
% \end{figure}

\section{Hyper parameters}

The ranker was trained for 3 epochs. Batch size of 64 sentences was used during training. Adam optimizer was used with a learning rate of 0.001. Ranker model consists of 4 linear layer, the first three layers having tanh activation and the final layer having a sigmoid activation function.

The Full training code is available here - \texttt{add github repo}

\section{Complete information about the datasets}
In this section we provide a complete overview about the datasets used for experiments :

We evaluate performance using the following datasets, each of which we split into ``sensitive'' and ``safe'' datasets.  
% \textbf{**add url links to the datasets}

(i) Medal dataset~\cite{medal}\footnote{\url{https://huggingface.co/datasets/medal}}. This dataset contains abstracts of medical papers, along with the diseases the abstract talks about. 
% \textbf{**say something about how this dataset was generated, what it contains, how it is labelled.} 
We partition this dataset into text with cancerous and non-cancerous diseases. Each dataset contains 2200 sentences. For our experiments, text with cancerous diseases was chosen to be the sensitive dataset. 
% \textbf{**presumably data is labelled to allow this?  what are sizes of the two datasets?}.  

(ii) Political dataset-\cite{politicsdata}\footnote{ Data can be downloaded by following the instructions in the repository
  \url{https://github.com/xuqiongkai/PATR}}. This contains comments on Facebook posts from 412 members of the United States Senate and House. Each comment is labeled with the corresponding Congressperson’s party affiliation i.e. S \(\epsilon\) \{democratic, republican \} 
% \textbf{**same comments: how was the dataset generated, what does it contain, how is it labelled}.  
We partition the dataset into text from users with Republican and Democrat political preferences. Each dataset contains 2000 sentences. For our experiments, text from users with Republican political preferences is  chosen to be the sensitive dataset. 
% \textbf{**presumably data is labelled to allow this?  what are sizes of the two datasets?  since we have sensitive and safe datasets, which one is safe here?}.  

(iii) {Amazon dataset%~\cite{redditsuicide} 
\footnote{
  \url{https://huggingface.co/datasets/amazon_reviews_multi}}.  
  This dataset contains product reviews from Amazon customers. We selected the reviews which were categorised as "drug-store" and "kitchen-appliances".
% \textbf{**say something about how this dataset was generated, what it contains, how it is labelled.}  
For our experiments, the dataset with drug-store reviews was chosen to be the sensitive dataset.}

(iv) Reddit dataset%~\cite{redditsuicide} 
\footnote{
  \url{https://www.kaggle.com/general/256134}}.  This dataset contains post content from the subreddits r/depression and r/SuicideWatch. We partition this data into posts related to suicide and depression. Each dataset contains 2000 sentences.
% \textbf{**say something about how this dataset was generated, what it contains, how it is labelled.}  
For our experiments, the text from the suicide subreddit was chosen to be the sensitive dataset.

% \section{KL-Divergence loss}
% We used KL-divergence loss for training purposes but used renyi divergence for validation purposes. Renyi-divergence had more non-differentiable operations hence we could not convert it to a loss function. We leave this for future work.

\begin{footnotesize}
\bibliographystyle{plain} % We choose the "plain" reference style
% \bibliography{refs}
\bibliography{bibliography}

\end{footnotesize}

% ****************************************************************************
% END OF BIBLIOGRAPHY AREA
% ****************************************************************************